\documentclass[manuscript, sigconf]{acmart}
\usepackage{hyperref}
\usepackage{enumerate}

\AtBeginDocument{%
  \providecommand\BibTeX{{%
    \normalfont B\kern-0.5em{\scshape i\kern-0.25em b}\kern-0.8em\TeX}}}

\setcopyright{rightsretained}
\copyrightyear{2024}
\acmYear{2024}
\acmDOI{3678884.3681829}
\acmConference[CSCW Companion '24]{Companion of the 2024 Computer-Supported Cooperative Work and Social Computing}{November 9--13, 2024}{San Jose, Costa Rica}
\acmBooktitle{Companion of the 2024 Computer-Supported Cooperative Work and Social Computing (CSCW Companion '24), November 9--13, 2024, San Jose, Costa Rica}\acmDOI{10.1145/3678884.3681829}
\acmISBN{979-8-4007-1114-5/24/11}

\begin{document}
\newcommand\angie[1]{{\color{purple} #1}}

%%
%% The "title" command has an optional parameter,
%% allowing the author to define a "short title" to be used in page headers.
\title{Worker Data Collectives as a means to Improve Accountability, Combat Surveillance and Reduce Inequalities }

%%
%% The "author" command and its associated commands are used to define
%% the authors and their affiliations.
%% Of note is the shared affiliation of the first two authors, and the
%% "authornote" and "authornotemark" commands
%% used to denote shared contribution to the research.
\author{Jane Hsieh}
\affiliation{%
  \institution{Carnegie Mellon University}
  \country{USA}
}

\author{Angie Zhang}
\affiliation{%
  \institution{University of Texas at Austin}
  \country{USA}
  }

\author{Seyun Kim }
\affiliation{%
  \institution{Carnegie Mellon University}
  \country{USA}
  }

\author{Varun Nagaraj Rao}
\affiliation{%
  \institution{Princeton University}
  \country{USA}
}

\author{Samantha Dalal}
\affiliation{%
  \institution{University of Colorado Boulder}
  \country{USA}
}

\author{Alexandra Mateescu}
\affiliation{
\institution{Data \& Society}
\country{USA}
}

\author{Rafael Do Nascimento Grohmann} 
\affiliation{
    \institution{University of Toronto Scarborough}
    \country{Canada}
}

\author{Motahhare Eslami}
\affiliation{%
 \institution{Carnegie Mellon University}
 \country{USA}}

\author{Min Kyung Lee}
\affiliation{%
  \institution{University of Texas at Austin}
  \country{USA}}

\author{Haiyi Zhu}
\affiliation{%
  \institution{Carnegie Mellon University}
  \country{USA}
}

%%
%% By default, the full list of authors will be used in the page
%% headers. Often, this list is too long, and will overlap
%% other information printed in the page headers. This command allows
%% the author to define a more concise list
%% of authors' names for this purpose.
\renewcommand{\shortauthors}{}

%%
%% The abstract is a short summary of the work to be presented in the
%% article.
\begin{abstract}
Platform-based laborers face unprecedented challenges and working conditions that result from algorithmic opacity, insufficient data transparency, and unclear policies and regulations. The CSCW and HCI communities increasingly turn to worker data collectives as a means to advance related policy and regulation, hold platforms accountable for data transparency and disclosure, and empower the collective worker voice. However, fundamental questions remain for designing, governing and sustaining such data infrastructures. In this workshop, we leverage frameworks such as data feminism to design sustainable and power-aware data collectives that tackle challenges present in various types of online labor platforms (e.g., ridesharing, freelancing, crowdwork, carework). While data collectives aim to support worker collectives and complement relevant policy initiatives, the goal of this workshop is to encourage their designers to consider topics of governance, privacy, trust, and transparency. In this one-day session, we convene research and advocacy community members to reflect on critical platform work issues (e.g., worker surveillance, discrimination, wage theft, insufficient platform accountability) as well as to collaborate on codesigning data collectives that ethically and equitably address these concerns by supporting working collectivism and informing policy development.
% \begin{itemize}
% \item what are some outstanding challenges and limitations tho around creating/designing worker data collectives and the potential they can provide: leveraging data feminism questions around how to create a sustainable platform for workers, balance power dynamics of different stakeholders, design it in complement of policy initiatives (thinking about the HCI and Policy workshop and paper qian yang et al. put out in CHIi this year) 
%     \item Goal of the workshop: we seek to convene the research and advocacy communities to codesign data collectives that help address issues present in different forms of platform work while accounting for challenges such as governance, privacy, trust and transparency; contextualize worker; ideate and exchange 
%     \item recruit participants who have interest in advancing the rights of workers from backgrounds/expertise of law, policy, worker-organizing, critical data studies, etc

% \end{itemize}

\end{abstract}

%%
%% The code below is generated by the tool at http://dl.acm.org/ccs.cfm.
%% Please copy and paste the code instead of the example below.
%%

%%
%% Keywords. The author(s) should pick words that accurately describe
%% the work being presented. Separate the keywords with commas.

%% A "teaser" image appears between the author and affiliation
%% information and the body of the document, and typically spans the
%% page.

%\received{20 February 2007}
%\received[revised]{12 March 2009}
%\received[accepted]{5 June 2009}

%%
%% This command processes the author and affiliation and title
%% information and builds the first part of the formatted document.
\maketitle

% \textcolor{red}{Add Infographic/overview}

\section{Introduction}
The emergence of platform-based work over the past decade disrupted labor markets in the US and across the globe. As of September 2023, the gig workforce was estimated to range from 154 to 435 million workers, comprising 4-13\% of the global labor force \footnote{The lower bound of 154 million or 4.4\% represents an estimate of only main/full-time workers whereas the upper bound of 435 million or 12.5\% also includes part-time/secondary workers } \cite{woborders}. Workers increasingly engage in platform-based gig work for a variety of reasons, including the promise of work flexibility and autonomy \cite{goodgig}, potential to mitigate discrimination (e.g., gender, ethnicity, sexuality, disability) as enabled by worker anonymity on certain platforms \cite{ghostwork} and opportunity for upskilling provided by macrotask or freelancing on-demand platforms \cite{ghostwork}.

But as platform-based labor (ridesharing/delivery drivers, caretakers, crowdworkers, capital platform workers, etc.) emerges as a flexible alternative to traditional employment, gig workers face unprecedented challenges and data harms \cite{redden2017data}: algorithmically-reinforced inequality and power differentials \cite{actor_network, goodgig, organizing, brown2021femtech}, overexposure to workplace monitoring and surveillance \cite{monitoring_explainer, explainer}, physical risks \cite{necrocapitalism, health_risks, technostress}, heightened uncertainty \cite{lauren12exploring, ball2021electronic}, and social isolation from peers \cite{atomized, wood2019networked}. 
Numerous nations intend to increase regulation of labor platforms \cite{us_regulate, international_regulation, australia_regulation}, but are limited by the scarcity of publicly accessible worker data  \cite{regulatory_data_need}. 

In resistance to surveillance and hegemonic data practices of platforms \cite{hegemonic, boss, privacy}, workers increasingly engage in self-tracking through individual means \cite{self_track} or third-party tools \footnote{e.g. \href{https://gridwise.io/}{Gridwise}, \href{https://www.stridehealth.com/tax}{Stride} and \href{https://www.strava.com/}{Strava}}. In the absence of sufficient policy and regulations for responsible platform practices, researchers and advocates increasingly turn to data collectives and tools as a method for advancing regulation \cite{organizing, supporting}, restoring worker power \cite{probes, sousveillance, uuapp, codesign} and holding platforms accountable to more ethical, fair and community-centered data practices\footnote{e.g., \href{https://getfairfare.org}{FairFare}, a worker auditing tool to uncover platform commission, and \href{https://www.driversseat.co/}{Driver's Seat Cooperative}, now under the \href{https://wao.cs.princeton.edu/}{Worker's Algorithm Observatory}, to help drivers and researchers investigate gig platform transparency and workers' experiences} \cite{explainer}. 

To define worker data collectives, we turn the HCI/CSCW literature for aggregating potential future data infrastructures that empower workers with informational and social support \cite{imaginaries, uuapp, codesign}. 
Recent efforts leveraged participatory design with workers and relevant stakeholders to reveal several (counter-)\textbf{data collectives} for supporting workers.
Such collective data institutions included 
% platform-initiated changes (e.g., feedback mechanisms, financial planning assistance \cite{uuapp, codesign}), 
online social institutions (e.g., collective wikis, social media groups/unions \cite{atomized}), 
offline social institutions (e.g., union strikes leveraging the power of social media to coalesce and organize \cite{la_strikes}),
third-party tools designed for data sharing \cite{uuapp, codesign, probes}, tracking \cite{, self_track, sousveillance}, and platform-evaluation (e.g., Fairwork \cite{fairwork}). 
% as well as apps that workers repurposed for self-tracking (e.g., Strava, Stride).
Regardless of the specific infrastructure, data collectives (by gathering and uplifting worker input on platformic work conditions) hold considerable promise for facilitating worker advocacy and empowerment, since they embody a site for communities of resistance \cite{critique} and enable collective forms of worker data actions such as counter-data collection and data refusal/strikes \cite{leverage, refusal, strikes}. 

For workers to fully enact the potential of data collectives as a vehicle for the production of counter-data and the restitution of power/rights over data, designers and maintainers must prioritize principles of care~\cite{data_care, sousveillance, d2022feminicide}, ethics \cite{fem_ethics} and justice \cite{justice, data_feminism_advocacy}.  We draw from seven principles of the intersectional feminist framework by D’Ignazio and Klein \cite{d2023data} and insights around workers' challenges informed by prior empirical work \cite{codesign, supporting} to consider ways of: 

\begin{itemize}
    \item Articulating invisible/unpaid work and addressing wage theft---\textit{Principle 7: Making Labor Visible} 
    \item Collectively auditing/disaggregating worker data withheld by platforms and challenging resultant algorithmic decisions---\textit{Principles 1 \& 2: Examining \& Challenging Power} 
    \item Addressing (physical and digital) safety risks that platforms fail to account for, including dangers present on roads, in strangers' homes, and from online scams---\textit{Principles 3 \& 6: Elevating Emotion and Embodiment by Considering Context}
    \item Gathering qualitative accounts/narratives of discrimination against marginalized individuals and work strategies---\textit{Principles 4 \& 5: Rethink Binaries and Hierarchies, Embrace Pluralism}
    \item Building infrastructure around interpreting and operationalizing assets in data collectives to precipitate material change---\textit{Principle 6: Considering Context}
\end{itemize}

Ultimately, advocates leveraging data collectives and tools aim to improve current labor regulations surrounding gig work data/conditions, or propose new litigation to advance worker (data) protections. 
To ensure that a policy-influencing data collective maintains long-term trust with workers, designers must consider effective ways of balancing governance and power structures as well as privacy protections, while allowing non-worker stakeholders to access necessary (group-level) data insights to make informed advocacy and policy decisions. 
Thus, in addition to considering methods of empowering workers with ethical, caring and just data practices, we aim to discuss in this workshop effective ways to design data collectives so they can serve as boundary objects that span across different stakeholders' needs and ways of knowing and collaborating. In this workshop, we refer to three types of stakeholders: workers, researchers, and practitioners (which include advocates, activist groups, lawmakers and policymakers, etc.).

%%Angie--I haven't read the full intro, but making sure we are emphasizing as motivation all the different research directions + organizer activity + policy efforts and how they're converging on the same theme? ---how to use data to support worker advocacy and empowerment?

% --->jake's work; data probes past work; reports by worker advocacy groups and their use of data; --> all converging on how to use worker's work data and narratives
% existing text encapsulates advocacy directions but not necessarily policy efforts yet

% part 1, already there--how to use data as a form of empowerment for workers 

% part 2, not there--motivation for investigating how to address issues of governance and privacy \cite{privacy} and trust (that workers trust the data collective and will contribute) and (long-term) sustainability  

% fold to under workshop goals
% \subsection{Guiding Questions}

%how can we 

% Paragraph summary: how to design data sharing ecosystems to address these practices/shortcomings, for example: trust, governance, privacy

%\section{Workshop Themes}
\section{Workshop Goals}
% The primary objectives of this workshop are to 1) share challenges and opportunities that researchers and civil society members face when advocating for worker data rights or related regulation and 2) identify how worker data collectives can serve as boundary objects that mediate meaningful collaborations amongst worker and non-worker stakeholders. %between researchers, advocates, policymakers and workers. 
% so as to facilitate collective efforts that tackle the issues of opaque algorithmic management and advance digital labor rights in practice. 
% In this session, we invite experts from diverse roles (researchers, advocates and a public policy analyst) and multiple nations to discuss, critique and explore the role that data-sharing collectives may play in advancing data and labor regulations as well as empowering algorithmically-impacted regions and populations of workers. 
% Below we outline concrete workshop goals, accompanied by relevant guiding questions:

The workshop goals and relevant guiding questions are as follows:

\begin{itemize}
    \item \textbf{Convene} a community of different stakeholder groups to discuss challenges and opportunities of worker data-sharing collectives for empowering platform workers. Many researcher, advocacy, and worker-organizing efforts have converged on the importance and necessity of worker data (practices) for auditing platforms, surfacing platform manipulation, or informing the need for policy and regulation \cite{uuapp, probes, homecarewages, datagovernance, blackbox}.
    This workshop will serve as an avenue for collaboration among these existing efforts.

    \begin{itemize}
        \item Whose preferences should the collective prioritize when gathering and analyzing worker data? 
        \item Which relevant stakeholder groups should participate in making changes to labor policy using data collectives?
        \item How can/should we design, deploy and sustain ecosystems of worker data collectives that benefit platform-based workers while respecting priorities and needs of related stakeholder groups?
    \end{itemize}
    
    \item \textbf{Contextualize} worker data within broader questions of worker rights, well-being and autonomy, including asking what kinds of worker data are meaningful, where data is shaped by conditions of constant worker surveillance, and the limitations of data as a tool.
    \begin{itemize}
    \item What are potential mechanisms for gathering and contextualizing novel worker data so as to de-invisibilize labor?% through uncovering previously unseen counter-data? --note from angie: i don't understand this part of the question so i commentedit out 
    \item Which data infrastructures can advance counter-data practices and policy changes that address pressing issues such as platform surveillance, discrimination and wage theft?
    \end{itemize} 
    
    \item \textbf{Ideate and exchange} perspectives on how such technologies can be governed and impact labor regulation across geographic regions/nations. In addition to constructing a shared understanding of the landscape, we aim to form a future research agenda.
    \begin{itemize}
        \item What are design and ethical considerations to keep in mind beyond protecting worker privacy, maintaining trust among stakeholders, and establishing reliable governance mechanisms?
        \item How can worker data collectives be designed to complement policy initiatives, e.g., enforcing platform data disclosures? In general, what role can/should policy play in supporting platform-based workers? %cite CO senate bill and washington state house bill---thanks Varun!!
    \end{itemize}
\end{itemize}

\section{Workshop Agenda \& Activities}\label{activities}
%%%%%Should we suggest different activities for online/in-person?%%%%%%% I haven't seen a workshop that separates online/in-person sessions, but happy to revisit this if we find one!
A tentative workshop schedule is outlined in Table \ref{table:agenda}. We will begin with a welcome keynote by 1-2 speaker(s) experienced in worker advocacy or labor policy.
% relevant worker organization
Next, participants will introduce their backgrounds and interests through lightning talks. 
Following a break, participants will engage in interactive group design and discussion to document ideas, themes, experiences, challenges/questions, and resources related to worker data collectives. Afterwards, each group will present the outcomes of their design. The workshop will conclude with a synthesis of high-level themes surfaced from presentations and a discussion of future directions.% and resolve lingering questions.
% \begin{table}[H]
% \begin{tabular}{|c|l|}
% \hline
% Time & \multicolumn{1}{c|}{Activity} \\ \hline
% 9-9:15am     & Welcome \& Icebreakers \\ \hline
% 9:15-10:30am & \begin{tabular}[c]{@{}l@{}}Activity 1: \\ Lightning Talk Introductions \& Reflections \end{tabular} \\ \hline
% 10:30-10:45am   & Coffee break \\ \hline
% 10:45-12:15pm   & \begin{tabular}[c]{@{}l@{}}Activity 2: \\ Co-Designing Worker Data Collectives \end{tabular} \\ \hline
% 12:15-1:30pm & Lunch \\ \hline
% 1:30-2:30pm  & \begin{tabular}[c]{@{}l@{}}Activity 3: \\ Presentation \& Artefact Share-Out \end{tabular} \\ \hline
% 2:30-3pm     & Coffee break \\ \hline
% 3-4:30pm     & \begin{tabular}[c]{@{}l@{}}Activity 4: \\ Takeaways and Future Directions \end{tabular}                                  \\ \hline
% 4:30-5pm     & Closing remarks \\ \hline
% \end{tabular}
% \caption{Agenda for the Workshop}
% \label{table:agenda}
% \end{table}

\begin{table}[H]
\begin{tabular}{ll}
\hline
\multicolumn{1}{c}{\textbf{Time}} & \multicolumn{1}{c}{\textbf{Activity}}       \\ \hline
9-9:30am                          & Welcome \& Keynote                          \\
9:30-10:30am                      & Lightning Talk Introductions \& Reflections \\ \hline
10:30-10:45am                     & \textit{Coffee break}                       \\ \hline
10:45am-11:30pm                   & Confronting Design Challenges               \\
11:30am-12:30pm                   & Co-Designing Worker Data Collectives        \\ \hline
12:30-1:30pm                      & \textit{Lunch}                              \\ \hline
1:30-2:45pm                       & Presentation \& Artefact Share-Out          \\ \hline
2:45-3pm                          & \textit{Coffee break}                       \\ \hline
3-4:30pm                          & Takeaways and Future Directions             \\ \hline
4:30-5pm                          & Closing remarks                            \\ \hline
\end{tabular}
\caption{Proposed Workshop Agenda}
\label{table:agenda}
\end{table}

\setcounter{subsection}{-1}

\subsection{Welcome \& Keynote}
%Old: To begin the workshop, a keynote speaker with practical experience at/with (non-profit) worker organizations will share their insights on challenges and opportunities related to labor advocacy for platform workers. We intend to invite either a practitioner with extensive experience advocating for platform workers or a leading academic researchers in the space.
To begin the workshop, 1-2 keynote speaker(s) with firsthand experience at/with (non-profit) worker organizations will share insights on challenges and opportunities related to labor advocacy for platform workers. We will extend the invitation to active worker-organizations (e.g., Rideshare Drivers United and Colorado Independent Drivers Union),
% who can speak to worker initiatives and practicalities of data efforts and needs
non-profit institutions (e.g., Colorado Fiscal Institute),
% who can describe data needs of the policy landscape; 
legal advocacy groups (e.g., Towards Justice),
% for perspectives about worker data for litigation; 
and leading academic researchers.
% to share their thoughts on this landscape of platform work and data collectives. %Several of the organizers have worked directly with groups or individuals in roles as mentioned above or would fall into those categories themselves (e.g., Rafael Do Nascimento Grohmann is a current member of Fairwork)

% Potential speakers we have been in conversation with include Wilneida Negrón from Coworker.org or members of Fairwork --- Rafael Do Nascimento Grohmann (organizing member).

\subsection{Activity 1: Lightning Talk Introductions \& Reflections} \label{activity1}
Participants will introduce themselves and share reflections on a question below as addressed in their submissions. Listening participants will be encouraged to respond with further reflections.

\begin{enumerate}
    \item How do we define power and who holds power among existing labor platform structures? 
    \item What are our roles (as non-worker stakeholders) when examining/designing data collectives?
    \item Whose goals are prioritized when analyzing and collecting data?
    \item What (forms of) counter-data should be collected and analyzed to challenge power?
    \item What role can disaggregated data (data divided into sub-categories) play in challenging power? 
\end{enumerate}

\subsection{Activity 2: Confronting Design Challenges}
\begin{figure}[h]
\centering
\includegraphics[width=9cm]{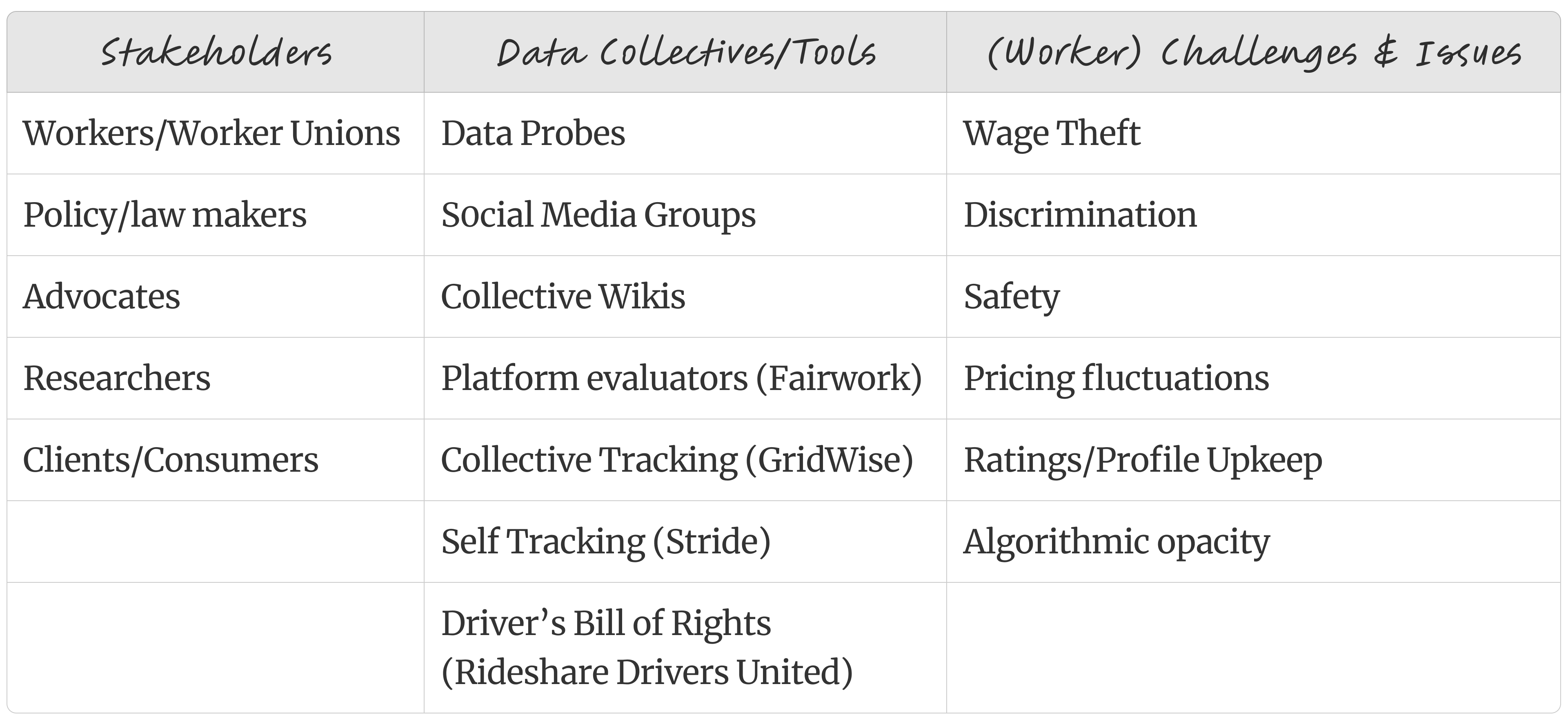}
\caption{Overview of Impacted Stakeholders, Worker Challenges/Issues and Potential Data Collectives}
\label{overview}
\end{figure}

Participants will brainstorm potential challenges in designing a worker data collective and issues of current platform work conditions for the system to address. Below are higher-level questions around data collective design and an overview of potential stakeholders, issues and data structures (Fig. \ref{overview}) to kick-start the session.

\begin{enumerate}
    \item What \textit{governance} and \textit{ownership} principles should designers of data collectives consider to ensure \textit{sustainability} \cite{datasharing_governance, buhler2023unlocking}?
    \item How do we build and maintain \textit{trust} with workers---core data contributors---in the data collective while simultaneously tackling their envisioned policies and futures? \cite{harrington2019deconstructing, sorries2024advocating}
    \item How do we ensure workers' \textit{privacy} while enabling non-worker stakeholders to use aggregate insights of the data collective \cite{king2023privacy, sannon2022privacy}? % without compromising equity
    \item What \textit{alternative methodologies} for designing future data collectives and surfacing diverse perspectives? For example, how might we use decolonizing methodologies to elicit non-western centric perspectives around privacy preferences and harms \cite{arora2019decolonizing}?
    \item What are unexpected trade-offs data collectives could present (e.g., the potential for data transparency to abstract out the worker and the need for qualitative insights) \cite{ananny2018seeing, zalnieriute2021transparency}?
\end{enumerate}

\subsection{Activity 3: Co-Designing Worker Data Collectives} \label{activity2}
Participants will break into groups. Each group will design data collective structure(s) for a specific platform/work type using digital templates (e.g., guided Miro boards) and/or physical materials (e.g., posters, sticky notes, markers). 
Participants' submissions will inform their group assignments.
% Group member assignments and work platforms addressed will be determined based on the platforms represented across participants' statements of interest. 
Examples of possible platform groupings include: 1) \textit{Rideshare \& Delivery} (e.g., Uber, Doordash), 2) \textit{Freelancing \& Macrotasking} (e.g., Upwork, Fiverr), 3) \textit{Microtasking} (e.g., Amazon Mechanical Turk, Crowdflower, Appen), 4) Caretaking and Household Work (e.g., Care.com, CareRev). When designing worker data collectives, we encourage participants to consider the following questions:

%these are relevant to context, invisible labor
\begin{enumerate}
    \item What labor is currently invisible? What forms of quantitative \textit{and} qualitative data can/should be collected to make visible this labor and combat oppressive labor practices?
    \item What are potential mechanisms for collecting data about invisible labor? What burdens and benefits might these present to workers?
    \item How should this data be presented to different stakeholders? %we don't have to be explicit about data visualizations 
    %\item This is related to the above How do we design data input modalities to facilitate easier articulation of unseen labor?
    \item Beyond platform-collected worker data, what are additional sources for pooling worker data? Potential examples include personal health informatics devices for geo-location data, photographs for sensemaking and reflection.
    \item What questions from workers (and non-worker stakeholders) can/should a data collective support/answer?
\end{enumerate}

\subsection{Activity 4: Presentation \& Artefact Share-Out}\label{activity3}
Each group will present their data collective from activity 3. This can include describing infrastructural decisions, ideas for addressing the design questions, and new concerns or questions that arose during discussions. Observing groups will be encouraged to ask follow-up questions and share reflections, while keeping in mind the questions below:

%these are relevant to pluralism---we need to revisit all these questions for relevancy to Activity 3. i.e., why ask these in Activity 3 not Activity 2?
\begin{enumerate}
    % \item How can different forms of data collection affect the situatedness of our understanding of worker struggles as represented through data? %angie's note: i don't understand this questions lol help me %jane: not sure either so combining with the next q!
    \item How do the choices on data sources and collection mechanisms impact the portrayal and situatedness of the highly subjective and contextualized worker experiences? 
    \item As groups that wield power over shaping the sociotechnical systems of gig platforms, how can non-worker stakeholders actively bring marginalized voices to the discussions of data collectives? 
    \item How do we ensure that \textit{all forms of labor} are made visible, not just privileging a small subset or a majority?
    \item What efforts should data collectives make to \textit{ensure trust} with marginalized communities? 
\end{enumerate}
    
\subsection{Activity 5: Discussion of Takeaways and Future Directions}\label{activity4}
%During group presentations, facilitators will document themes about the emerging opportunities and challenges to frame the final discussion activity. 
To frame the final discussion, facilitators will summarize opportunities and challenges based on participants' ideas, questions, and concerns. Participants will be given space to consider and propose future research agendas or avenues of work. 
% During this time, facilitators will encourage considerations for topics of trust, governance, privacy and sustainability, as detailed below.

% Topics of consideration during discussion can include:
% Principles 1/2: Examining/Challenging Power
%--governance, privacy, trust, sustainability 
%- decolonizing methodologies for privacy/(governance)
% discuss these earlier, before codesign

% Trust ---> Workers should be able to have insights that immediately benefit them, not just waiting for aggregate results and policy to pass 
% How to address governance and privacy \cite{privacy} and trust (that workers trust the data collective and will contribute), security and sustainability of a collective.

%other ideas, have small groups that rotate out people

\section{Post-Workshop Activities}
%Workshop artefacts (e.g., Miro board links) will remain accessible after the event. 
Post-workshop, a document will be shared to participants to summarize each group's designed data collective with a) a link to the correlating Miro board, b) photos of physical artefacts created if applicable, c) a summary of the group's presentation and questions surfaced by others, and d) questions and themes from the talk-back session. Furthermore, we seek to support continuing collaboration interests that arise---for example, we may create a shared document for participants to share new resources or set up a collaborative platform to facilitate cross-organizational efforts related advancing work data collectives. Inspired by the workshop by Yang et. al. on bridging HCI and policy design, we may also consider synthesizing workshop insights into a provocation/position paper.

\section{Logistics}
This workshop will run as a full-day hybrid workshop to allow participation from a diverse range of geographic locations and backgrounds. Sessions will be mediated through Zoom and asynchronous conversations will be facilitated via Slack . 

\subsection{Participant Recruitment \& Selection}  
We will recruit a maximum of 50 participants who work on or demonstrate interest in platform-based labor. %jane: modifying language from on-demand labor as it excludes some pre-scheduled taskers like freelancers and petsitters
This includes researchers with backgrounds in Computer-Supported Cooperative Work, Human-Computer Interaction, Public Policy, Law (and beyond), as well as organizers, activists, and platform workers. We will distribute the workshop call through mailing lists and social media to reach relevant individuals.
% researching or actively involved in platform-based work. %easier/more accessible for workers to attend online than in-person and their feedback is valuable 
\subsection{Submission Formats \& Requirements:}
To be considered, participants should provide a statement of interest, either as 1) a maximum 500-word personal statement describing their background and motivation/interest in attending the workshop; or 2) a maximum two-page extended abstract or case study about a specific type of platform-based work as related to the workshop themes---e.g., speculation about challenges and opportunities, description of related prior/on-going work. The statement should address the question: \textit{How data can inform policymaking}? To optimize group assignments, we recommend submissions specify the type(s) of platforms/work where they have the most interest/experience. %%3) a maximum two-page case study discussing a specific challenge or effort about supporting platform-based workers using shared data. %What are your current understandings around how data can inform policymaking

We highly encourage submissions to reflect on concepts of power, ethics and their own positionality as related to platform-based work and counter-data. The guiding questions of
\nameref{activity1} can provide a starting ground for such discussions.
% concepts of \textit{power} and \textit{counter-data}, particularly in the context of the guiding questions that will be discussed during  (see Section  for the questions).}
Submissions that incorporate figures and/or diagrams ideating data sharing structures are welcomed but not required; figures, diagrams, and references do not count towards the page limit. 
% Statements of accepted participants will be uploaded to the workshop website. 

%Angie 5-11-2024 --> reword the recruitment to require people's position papers to address specific questions 

% We will be inviting workers to participate through organizers’ contacts, and they will be encouraged (but not required) to share their experiences and opinions during the session.

% \subsection{Facilitation}
% Workshop organizers will facilitate both in-person and online sessions.
\subsection{Resources Required}
\textbf{Equipment and Supplies Needed to Run the Workshop:} 
To accommodate in-person participants, we request access to standard conference room facilities, including seating for up to 25 participants, A/V equipment, and access to physical design resources (e.g., markers, sticky-notes, posters/whiteboards/large easel pads).% for the co-design activity. 
\\
\textbf{Resources participants are expected to bring or provide:} 
Online participants will need access to a desktop computer or laptop with internet connectivity to participate. 
In-person participants will also be expected to bring laptops in order to participate in the Miro board activities, and optionally to access their own and/or other participants statements of interest. 

\section{Workshop Organizers}
\textbf{Jane Hsieh }is a Ph.D. student at CMU studying ways of supporting gig workers through policy and technology advancements. She investigates issues and solutions (e.g. data-sharing alternatives) of platform-based work from multi-stakeholder perspectives. \\
\textbf{Angie Zhang }is a Ph.D. student at UT Austin, where she uses co-design to understand impacts of technology and AI on individuals and communities, and develops frameworks and prototyping tools to support participatory AI design with diverse stakeholders. \\
\textbf{Seyun Kim} is a Ph.D. student in CMU researching how to operationalize equity and fairness notions in data-driven technologies, particularly in clinical and public-sector technologies by surfacing perceptions of stakeholders like workers and entrepreneurs. \\
\textbf{Varun Nagaraj Rao} is a Ph.D. student at Princeton and is affiliated with the Center for Information Technology Policy (CITP) and the Workers Algorithm Observatory (WAO). He studies the societal impacts of AI systems in the context of labor and seeks to understand and build tools to mitigate concerns and needs of rideshare workers impacted by non-transparent AI and algorithmic decisions. \\
\textbf{Samantha Dalal} is a Ph.D student at CU Boulder investigating how information science researchers can precipitate change in the gig economy by participating in grassroots organizing and creation of labor policy at the state level. \\
\textbf{Alexandra Mateescu} is a researcher at Data \& Society, where she investigated workers’ experiences with data-centric worker surveillance  and intersections with broader inequalities across several U.S. contexts, including domestic work on online labor platforms, care work within public benefits programs, retail, and other industries. \\
\textbf{Rafael Grohmann} is an Assistant Professor of Media Studies (Critical Platform Studies) at the University of Toronto Scarborough. He is the PI for Worker-Owned Intersectional Platforms (WOIP -- an action research with delivery and tech workers in Brazil and Argentina), an editor for the journal Platforms and Society, and a member of Fairwork and Tierra Común.   \\
\textbf{Motahhare Eslami} is an Assistant professor at CMU investigating accountability challenges in algorithmic systems, and aims to empower the users of algorithmic systems to make transparent, fair, and informed decisions. \\
\textbf{Min Kyung Lee} is an Assistant professor at UT Austin working on building just and empowering workplaces and cities via technology that strengthen individual and collective human decision-making. \\
\textbf{Haiyi Zhu} is an Associate Professor at CMU interested in incorporating stakeholder values and societal considerations into the creation and evaluation process of AI technologies. She conducted several projects studying the design and social impacts of AI in gig work, child welfare, online content moderation, and mental health.

% \section{Call for Participation}
% \subsection{Website}

%%
%% The next two lines define the bibliography style to be used, and
%% the bibliography file.
\bibliographystyle{ACM-Reference-Format}
\balance
\bibliography{references}

%%% -*-BibTeX-*-
%%% Do NOT edit. File created by BibTeX with style
%%% ACM-Reference-Format-Journals [18-Jan-2012].

\begin{thebibliography}{54}

%%% ====================================================================
%%% NOTE TO THE USER: you can override these defaults by providing
%%% customized versions of any of these macros before the \bibliography
%%% command.  Each of them MUST provide its own final punctuation,
%%% except for \shownote{}, \showDOI{}, and \showURL{}.  The latter two
%%% do not use final punctuation, in order to avoid confusing it with
%%% the Web address.
%%%
%%% To suppress output of a particular field, define its macro to expand
%%% to an empty string, or better, \unskip, like this:
%%%
%%% \newcommand{\showDOI}[1]{\unskip}   % LaTeX syntax
%%%
%%% \def \showDOI #1{\unskip}           % plain TeX syntax
%%%
%%% ====================================================================

\ifx \showCODEN    \undefined \def \showCODEN     #1{\unskip}     \fi
\ifx \showDOI      \undefined \def \showDOI       #1{#1}\fi
\ifx \showISBNx    \undefined \def \showISBNx     #1{\unskip}     \fi
\ifx \showISBNxiii \undefined \def \showISBNxiii  #1{\unskip}     \fi
\ifx \showISSN     \undefined \def \showISSN      #1{\unskip}     \fi
\ifx \showLCCN     \undefined \def \showLCCN      #1{\unskip}     \fi
\ifx \shownote     \undefined \def \shownote      #1{#1}          \fi
\ifx \showarticletitle \undefined \def \showarticletitle #1{#1}   \fi
\ifx \showURL      \undefined \def \showURL       {\relax}        \fi
% The following commands are used for tagged output and should be
% invisible to TeX
\providecommand\bibfield[2]{#2}
\providecommand\bibinfo[2]{#2}
\providecommand\natexlab[1]{#1}
\providecommand\showeprint[2][]{arXiv:#2}

\bibitem[Ananny and Crawford(2018)]%
        {ananny2018seeing}
\bibfield{author}{\bibinfo{person}{Mike Ananny} {and} \bibinfo{person}{Kate Crawford}.} \bibinfo{year}{2018}\natexlab{}.
\newblock \showarticletitle{Seeing without knowing: Limitations of the transparency ideal and its application to algorithmic accountability}.
\newblock \bibinfo{journal}{\emph{new media \& society}} \bibinfo{volume}{20}, \bibinfo{number}{3} (\bibinfo{year}{2018}), \bibinfo{pages}{973--989}.
\newblock
\urldef\tempurl%
\url{https://doi.org/10.1177/1461444816676645}
\showDOI{\tempurl}


\bibitem[Arora(2019)]%
        {arora2019decolonizing}
\bibfield{author}{\bibinfo{person}{Payal Arora}.} \bibinfo{year}{2019}\natexlab{}.
\newblock \showarticletitle{Decolonizing privacy studies}.
\newblock \bibinfo{journal}{\emph{Television \& New Media}} \bibinfo{volume}{20}, \bibinfo{number}{4} (\bibinfo{year}{2019}), \bibinfo{pages}{366--378}.
\newblock
\urldef\tempurl%
\url{https://doi.org/10.1177/1527476418806092}
\showDOI{\tempurl}


\bibitem[Asih et~al\mbox{.}(2022)]%
        {hegemonic}
\bibfield{author}{\bibinfo{person}{Irsanti~Widuri Asih}, \bibinfo{person}{Heru Nugroho}, {and} \bibinfo{person}{Budiawan Budiawan}.} \bibinfo{year}{2022}\natexlab{}.
\newblock \showarticletitle{Hegemonic dialectics between power and resistance in the Indonesian sharing economy: Study of Gojek}.
\newblock \bibinfo{journal}{\emph{Informasi}} \bibinfo{volume}{52}, \bibinfo{number}{1} (\bibinfo{year}{2022}), \bibinfo{pages}{63--82}.
\newblock
\urldef\tempurl%
\url{https://doi.org/10.21831/informasi.v52i1.49348}
\showDOI{\tempurl}


\bibitem[Bajwa et~al\mbox{.}(2018)]%
        {health_risks}
\bibfield{author}{\bibinfo{person}{Uttam Bajwa}, \bibinfo{person}{Denise Gastaldo}, \bibinfo{person}{Erica Di~Ruggiero}, {and} \bibinfo{person}{Lilian Knorr}.} \bibinfo{year}{2018}\natexlab{}.
\newblock \showarticletitle{The health of workers in the global gig economy}.
\newblock \bibinfo{journal}{\emph{Global. Health}} \bibinfo{volume}{14}, \bibinfo{number}{1} (\bibinfo{date}{Dec.} \bibinfo{year}{2018}), \bibinfo{pages}{124}.
\newblock
\urldef\tempurl%
\url{https://doi.org/10.1186/s12992-018-0444-8}
\showDOI{\tempurl}


\bibitem[Ball et~al\mbox{.}(2021)]%
        {ball2021electronic}
\bibfield{author}{\bibinfo{person}{Kirstie Ball} {et~al\mbox{.}}} \bibinfo{year}{2021}\natexlab{}.
\newblock \showarticletitle{Electronic monitoring and surveillance in the workplace}.
\newblock \bibinfo{journal}{\emph{European Commission Joint Research Centre}} (\bibinfo{year}{2021}).
\newblock
\urldef\tempurl%
\url{https://doi.org/10.2760/5137}
\showDOI{\tempurl}


\bibitem[Benjamin(2021)]%
        {critique}
\bibfield{author}{\bibinfo{person}{Garfield Benjamin}.} \bibinfo{year}{2021}\natexlab{}.
\newblock \showarticletitle{What we do with data: a performative critique of data “collection”}.
\newblock \bibinfo{journal}{\emph{Internet Policy Review}} \bibinfo{volume}{10}, \bibinfo{number}{4} (\bibinfo{year}{2021}).
\newblock
\showISSN{2197-6775}
\urldef\tempurl%
\url{https://doi.org/10.14763/2021.4.1588}
\showDOI{\tempurl}


\bibitem[Boone et~al\mbox{.}(2023)]%
        {data_care}
\bibfield{author}{\bibinfo{person}{Ashley Boone}, \bibinfo{person}{Carl Disalvo}, {and} \bibinfo{person}{Christopher~A Le~Dantec}.} \bibinfo{year}{2023}\natexlab{}.
\newblock \showarticletitle{Data Practice for a Politics of Care: Food Assistance as a Site of Careful Data Work}. In \bibinfo{booktitle}{\emph{Proceedings of the 2023 CHI Conference on Human Factors in Computing Systems}} (Hamburg, Germany) \emph{(\bibinfo{series}{CHI '23})}. \bibinfo{publisher}{Association for Computing Machinery}, \bibinfo{address}{New York, NY, USA}, Article \bibinfo{articleno}{152}, \bibinfo{numpages}{13}~pages.
\newblock
\showISBNx{9781450394215}
\urldef\tempurl%
\url{https://doi.org/10.1145/3544548.3580831}
\showDOI{\tempurl}


\bibitem[Brown(2021)]%
        {brown2021femtech}
\bibfield{author}{\bibinfo{person}{Elizabeth~A Brown}.} \bibinfo{year}{2021}\natexlab{}.
\newblock \showarticletitle{The FemTech paradox: How workplace monitoring threatens women’s equity}.
\newblock \bibinfo{journal}{\emph{Jurimetrics}} \bibinfo{volume}{61}, \bibinfo{number}{3} (\bibinfo{year}{2021}), \bibinfo{pages}{289--329}.
\newblock
\urldef\tempurl%
\url{https://www.proquest.com/scholarly-journals/femtech-paradox-how-workplace-monitoring/docview/2568314630/se-2}
\showURL{%
\tempurl}


\bibitem[B{\"u}hler et~al\mbox{.}(2023)]%
        {buhler2023unlocking}
\bibfield{author}{\bibinfo{person}{Michael~Max B{\"u}hler}, \bibinfo{person}{Igor Calzada}, \bibinfo{person}{Isabel Cane}, \bibinfo{person}{Thorsten Jelinek}, \bibinfo{person}{Astha Kapoor}, \bibinfo{person}{Morshed Mannan}, \bibinfo{person}{Sameer Mehta}, \bibinfo{person}{Vijay Mookerje}, \bibinfo{person}{Konrad N{\"u}bel}, \bibinfo{person}{Alex Pentland}, {et~al\mbox{.}}} \bibinfo{year}{2023}\natexlab{}.
\newblock \showarticletitle{Unlocking the power of digital commons: Data cooperatives as a pathway for data sovereign, innovative and equitable digital communities}.
\newblock \bibinfo{journal}{\emph{Digital}} \bibinfo{volume}{3}, \bibinfo{number}{3} (\bibinfo{year}{2023}), \bibinfo{pages}{146--171}.
\newblock
\urldef\tempurl%
\url{https://doi.org/10.3390/digital3030011}
\showDOI{\tempurl}


\bibitem[Calacci(2022)]%
        {organizing}
\bibfield{author}{\bibinfo{person}{Dana Calacci}.} \bibinfo{year}{2022}\natexlab{}.
\newblock \showarticletitle{Organizing in the End of Employment: Information Sharing, Data Stewardship, and Digital Workerism}. In \bibinfo{booktitle}{\emph{Proceedings of the 1st Annual Meeting of the Symposium on Human-Computer Interaction for Work}} (Durham, NH, USA) \emph{(\bibinfo{series}{CHIWORK '22})}. \bibinfo{publisher}{Association for Computing Machinery}, \bibinfo{address}{New York, NY, USA}, Article \bibinfo{articleno}{14}, \bibinfo{numpages}{9}~pages.
\newblock
\showISBNx{9781450396554}
\urldef\tempurl%
\url{https://doi.org/10.1145/3533406.3533424}
\showDOI{\tempurl}


\bibitem[Calacci and Pentland(2022)]%
        {blackbox}
\bibfield{author}{\bibinfo{person}{Dana Calacci} {and} \bibinfo{person}{Alex Pentland}.} \bibinfo{year}{2022}\natexlab{}.
\newblock \showarticletitle{Bargaining with the Black-Box: Designing and Deploying Worker-Centric Tools to Audit Algorithmic Management}.
\newblock \bibinfo{journal}{\emph{Proc. ACM Hum.-Comput. Interact.}} \bibinfo{volume}{6}, \bibinfo{number}{CSCW2}, Article \bibinfo{articleno}{428} (\bibinfo{date}{nov} \bibinfo{year}{2022}), \bibinfo{numpages}{24}~pages.
\newblock
\urldef\tempurl%
\url{https://doi.org/10.1145/3570601}
\showDOI{\tempurl}


\bibitem[Calacci and Stein(2023)]%
        {datagovernance}
\bibfield{author}{\bibinfo{person}{Dana Calacci} {and} \bibinfo{person}{Jake Stein}.} \bibinfo{year}{2023}\natexlab{}.
\newblock \showarticletitle{From access to understanding: Collective data governance for workers}.
\newblock \bibinfo{journal}{\emph{European Labour Law Journal}} \bibinfo{volume}{14}, \bibinfo{number}{2} (\bibinfo{year}{2023}), \bibinfo{pages}{253--282}.
\newblock
\urldef\tempurl%
\url{https://doi.org/10.1177/20319525231167981}
\showDOI{\tempurl}


\bibitem[Collier et~al\mbox{.}(2017)]%
        {us_regulate}
\bibfield{author}{\bibinfo{person}{Ruth~Berins Collier}, \bibinfo{person}{Veena Dubal}, {and} \bibinfo{person}{Christopher Carter}.} \bibinfo{year}{2017}\natexlab{}.
\newblock \showarticletitle{Labor platforms and gig work: the failure to regulate}.
\newblock  (\bibinfo{year}{2017}).
\newblock
\urldef\tempurl%
\url{https://doi.org/10.2139/ssrn.3039742}
\showDOI{\tempurl}


\bibitem[Cram et~al\mbox{.}(2022)]%
        {technostress}
\bibfield{author}{\bibinfo{person}{W~Alec Cram}, \bibinfo{person}{Martin Wiener}, \bibinfo{person}{Monideepa Tarafdar}, {and} \bibinfo{person}{Alexander Benlian}.} \bibinfo{year}{2022}\natexlab{}.
\newblock \showarticletitle{Examining the impact of algorithmic control on Uber drivers’ technostress}.
\newblock \bibinfo{journal}{\emph{Journal of management information systems}} \bibinfo{volume}{39}, \bibinfo{number}{2} (\bibinfo{year}{2022}), \bibinfo{pages}{426--453}.
\newblock
\urldef\tempurl%
\url{https://doi.org/10.1080/07421222.2022.2063556}
\showDOI{\tempurl}


\bibitem[Darian et~al\mbox{.}(2023)]%
        {data_feminism_advocacy}
\bibfield{author}{\bibinfo{person}{Shiva Darian}, \bibinfo{person}{Aarjav Chauhan}, \bibinfo{person}{Ricky Marton}, \bibinfo{person}{Janet Ruppert}, \bibinfo{person}{Kathleen Anderson}, \bibinfo{person}{Ryan Clune}, \bibinfo{person}{Madeline Cupchak}, \bibinfo{person}{Max Gannett}, \bibinfo{person}{Joel Holton}, \bibinfo{person}{Elizabeth Kamas}, \bibinfo{person}{Jason Kibozi-Yocka}, \bibinfo{person}{Devin Mauro-Gallegos}, \bibinfo{person}{Simon Naylor}, \bibinfo{person}{Meghan O'Malley}, \bibinfo{person}{Mehul Patel}, \bibinfo{person}{Jack Sandberg}, \bibinfo{person}{Troy Siegler}, \bibinfo{person}{Ryan Tate}, \bibinfo{person}{Abigil Temtim}, \bibinfo{person}{Samantha Whaley}, {and} \bibinfo{person}{Amy Voida}.} \bibinfo{year}{2023}\natexlab{}.
\newblock \showarticletitle{Enacting Data Feminism in Advocacy Data Work}.
\newblock \bibinfo{journal}{\emph{Proc. ACM Hum.-Comput. Interact.}} \bibinfo{volume}{7}, \bibinfo{number}{CSCW1}, Article \bibinfo{articleno}{47} (\bibinfo{date}{apr} \bibinfo{year}{2023}), \bibinfo{numpages}{28}~pages.
\newblock
\urldef\tempurl%
\url{https://doi.org/10.1145/3579480}
\showDOI{\tempurl}


\bibitem[Datta et~al\mbox{.}(2006)]%
        {woborders}
\bibfield{author}{\bibinfo{person}{Namita Datta}, \bibinfo{person}{Chen Rong}, \bibinfo{person}{Sunamika Singh}, \bibinfo{person}{Clara Stinshoff}, \bibinfo{person}{Nadina Iacob}, \bibinfo{person}{Natnael~Simachew Nigatu}, \bibinfo{person}{Mpumelelo Nxumalo}, {and} \bibinfo{person}{Luka Klimaviciute}.} \bibinfo{year}{2006}\natexlab{}.
\newblock \bibinfo{title}{Working Without Borders: The Promise and Peril of Online Gig Work}.
\newblock
\newblock
\urldef\tempurl%
\url{https://www.worldbank.org/en/topic/jobsanddevelopment/publication/online-gig-work-enabled-by-digital-platforms}
\showURL{%
\tempurl}


\bibitem[De~Stefano(2018)]%
        {international_regulation}
\bibfield{author}{\bibinfo{person}{Valerio De~Stefano}.} \bibinfo{year}{2018}\natexlab{}.
\newblock \showarticletitle{The gig economy and labour regulation: an international and comparative approach}.
\newblock \bibinfo{journal}{\emph{Law Journal of Social and Labor Relations}}  \bibinfo{volume}{4} (\bibinfo{year}{2018}), \bibinfo{pages}{68}.
\newblock
Issue 2.
\urldef\tempurl%
\url{https://doi.org/10.26843/mestradodireito.v4i2.158}
\showDOI{\tempurl}


\bibitem[D'Ignazio et~al\mbox{.}(2022)]%
        {d2022feminicide}
\bibfield{author}{\bibinfo{person}{Catherine D'Ignazio}, \bibinfo{person}{Isadora Crux{\^e}n}, \bibinfo{person}{Helena~Su{\'a}rez Val}, \bibinfo{person}{Angeles~Martinez Cuba}, \bibinfo{person}{Mariel Garc{\'\i}a-Montes}, \bibinfo{person}{Silvana Fumega}, \bibinfo{person}{Harini Suresh}, {and} \bibinfo{person}{Wonyoung So}.} \bibinfo{year}{2022}\natexlab{}.
\newblock \showarticletitle{Feminicide and counterdata production: Activist efforts to monitor and challenge gender-related violence}.
\newblock \bibinfo{journal}{\emph{Patterns}} \bibinfo{volume}{3}, \bibinfo{number}{7} (\bibinfo{year}{2022}).
\newblock
\urldef\tempurl%
\url{https://doi.org/10.1016/j.patter.2022.100530}
\showDOI{\tempurl}


\bibitem[D'ignazio and Klein(2023)]%
        {d2023data}
\bibfield{author}{\bibinfo{person}{Catherine D'ignazio} {and} \bibinfo{person}{Lauren~F Klein}.} \bibinfo{year}{2023}\natexlab{}.
\newblock \bibinfo{booktitle}{\emph{Data feminism}}.
\newblock \bibinfo{publisher}{MIT press}.
\newblock


\bibitem[Do et~al\mbox{.}(2024)]%
        {sousveillance}
\bibfield{author}{\bibinfo{person}{Kimberly Do}, \bibinfo{person}{Maya De~Los Santos}, \bibinfo{person}{Michael Muller}, {and} \bibinfo{person}{Saiph Savage}.} \bibinfo{year}{2024}\natexlab{}.
\newblock \showarticletitle{Designing Sousveillance Tools for Gig Workers}. In \bibinfo{booktitle}{\emph{Proceedings of the 2023 CHI Conference on Human Factors in Computing Systems}} \emph{(\bibinfo{series}{CHI '24})}. \bibinfo{publisher}{Association for Computing Machinery}, \bibinfo{address}{New York, NY, USA}.
\newblock
\urldef\tempurl%
\url{https://doi.org/10.1145/3613904.3642614}
\showDOI{\tempurl}


\bibitem[Draude et~al\mbox{.}(2022)]%
        {justice}
\bibfield{author}{\bibinfo{person}{Claude Draude}, \bibinfo{person}{Gerrit Hornung}, {and} \bibinfo{person}{Goda Klumbyt{\.e}}.} \bibinfo{year}{2022}\natexlab{}.
\newblock \showarticletitle{Mapping data justice as a multidimensional concept through feminist and legal perspectives}.
\newblock In \bibinfo{booktitle}{\emph{New Perspectives in Critical Data Studies: The Ambivalences of Data Power}}. \bibinfo{publisher}{Springer International Publishing Cham}, \bibinfo{pages}{187--216}.
\newblock
\urldef\tempurl%
\url{https://doi.org/10.1007/978-3-030-96180-0_9}
\showDOI{\tempurl}


\bibitem[Eckartz et~al\mbox{.}(2014)]%
        {datasharing_governance}
\bibfield{author}{\bibinfo{person}{Silja~M Eckartz}, \bibinfo{person}{Wout~J Hofman}, {and} \bibinfo{person}{Anne~Fleur Van~Veenstra}.} \bibinfo{year}{2014}\natexlab{}.
\newblock \showarticletitle{A decision model for data sharing}. In \bibinfo{booktitle}{\emph{Electronic Government: 13th IFIP WG 8.5 International Conference, EGOV 2014, Dublin, Ireland, September 1-3, 2014. Proceedings 13}}. Springer, \bibinfo{pages}{253--264}.
\newblock
\urldef\tempurl%
\url{https://doi.org/10.1007/978-3-662-44426-9_21}
\showDOI{\tempurl}


\bibitem[Graham and Woodcock(2018)]%
        {fairwork}
\bibfield{author}{\bibinfo{person}{Mark Graham} {and} \bibinfo{person}{Jamie Woodcock}.} \bibinfo{year}{2018}\natexlab{}.
\newblock \showarticletitle{Towards a fairer platform economy: introducing the Fairwork Foundation}.
\newblock \bibinfo{journal}{\emph{Alternate Routes}}  \bibinfo{volume}{29} (\bibinfo{year}{2018}).
\newblock
\urldef\tempurl%
\url{https://alternateroutes.ca/index.php/ar/article/view/22455}
\showURL{%
\tempurl}


\bibitem[Gray and Suri(2019)]%
        {ghostwork}
\bibfield{author}{\bibinfo{person}{Mary~L Gray} {and} \bibinfo{person}{Siddharth Suri}.} \bibinfo{year}{2019}\natexlab{}.
\newblock \bibinfo{booktitle}{\emph{Ghost work: How to stop Silicon Valley from building a new global underclass}}.
\newblock \bibinfo{publisher}{Eamon Dolan Books}.
\newblock


\bibitem[Harrington et~al\mbox{.}(2019)]%
        {harrington2019deconstructing}
\bibfield{author}{\bibinfo{person}{Christina Harrington}, \bibinfo{person}{Sheena Erete}, {and} \bibinfo{person}{Anne~Marie Piper}.} \bibinfo{year}{2019}\natexlab{}.
\newblock \showarticletitle{Deconstructing Community-Based Collaborative Design: Towards More Equitable Participatory Design Engagements}.
\newblock \bibinfo{journal}{\emph{Proc. ACM Hum.-Comput. Interact.}} \bibinfo{volume}{3}, \bibinfo{number}{CSCW}, Article \bibinfo{articleno}{216} (\bibinfo{date}{nov} \bibinfo{year}{2019}), \bibinfo{numpages}{25}~pages.
\newblock
\urldef\tempurl%
\url{https://doi.org/10.1145/3359318}
\showDOI{\tempurl}


\bibitem[Hawley(2018)]%
        {regulatory_data_need}
\bibfield{author}{\bibinfo{person}{Adrian~John Hawley}.} \bibinfo{year}{2018}\natexlab{}.
\newblock \showarticletitle{Regulating labour platforms, the data deficit}.
\newblock \bibinfo{journal}{\emph{European Journal of Government and Economics}} \bibinfo{volume}{7}, \bibinfo{number}{1} (\bibinfo{year}{2018}), \bibinfo{pages}{5--23}.
\newblock
\urldef\tempurl%
\url{https://doi.org/10.17979/ejge.2018.7.1.4330%0A}
\showDOI{\tempurl}


\bibitem[Hernandez et~al\mbox{.}(2024)]%
        {self_track}
\bibfield{author}{\bibinfo{person}{Rie Helene~(Lindy) Hernandez}, \bibinfo{person}{Qiurong Song}, \bibinfo{person}{Yubo Kou}, {and} \bibinfo{person}{Xinning Gui}.} \bibinfo{year}{2024}\natexlab{}.
\newblock \showarticletitle{"At the end of the day, I am accountable": Gig Workers' Self-Tracking for Multi-Dimensional Accountability Management}. In \bibinfo{booktitle}{\emph{Proceedings of the CHI Conference on Human Factors in Computing Systems}} (Honolulu, HI, USA) \emph{(\bibinfo{series}{CHI '24})}. \bibinfo{publisher}{Association for Computing Machinery}, \bibinfo{address}{New York, NY, USA}, Article \bibinfo{articleno}{382}, \bibinfo{numpages}{20}~pages.
\newblock
\showISBNx{9798400703300}
\urldef\tempurl%
\url{https://doi.org/10.1145/3613904.3642151}
\showDOI{\tempurl}


\bibitem[Howson et~al\mbox{.}(2020)]%
        {la_strikes}
\bibfield{author}{\bibinfo{person}{Kelle Howson}, \bibinfo{person}{Funda Ustek-Spilda}, \bibinfo{person}{Rafael Grohmann}, \bibinfo{person}{Nancy Salem}, \bibinfo{person}{Rodrigo Carelli}, \bibinfo{person}{Daniel Abs}, \bibinfo{person}{Julice Salvagni}, \bibinfo{person}{Mark Graham}, \bibinfo{person}{Maria~Belen Balbornoz}, \bibinfo{person}{Henry Chavez}, {et~al\mbox{.}}} \bibinfo{year}{2020}\natexlab{}.
\newblock \showarticletitle{`Just because you don't see your boss, doesn't mean you don't have a boss`: Covid-19 and Gig Worker Strikes across Latin America}.
\newblock \bibinfo{journal}{\emph{International Union Rights}} \bibinfo{volume}{27}, \bibinfo{number}{3} (\bibinfo{year}{2020}), \bibinfo{pages}{20--28}.
\newblock
\urldef\tempurl%
\url{https://doi.org/10.1353/iur.2020.a838172}
\showDOI{\tempurl}


\bibitem[Hsieh et~al\mbox{.}(2023)]%
        {codesign}
\bibfield{author}{\bibinfo{person}{Jane Hsieh}, \bibinfo{person}{Miranda Karger}, \bibinfo{person}{Lucas Zagal}, {and} \bibinfo{person}{Haiyi Zhu}.} \bibinfo{year}{2023}\natexlab{}.
\newblock \showarticletitle{Co-Designing Alternatives for the Future of Gig Worker Well-Being: Navigating Multi-Stakeholder Incentives and Preferences}. In \bibinfo{booktitle}{\emph{Proceedings of the 2023 ACM Designing Interactive Systems Conference}} (Pittsburgh, PA, USA,) \emph{(\bibinfo{series}{DIS '23})}. \bibinfo{publisher}{Association for Computing Machinery}, \bibinfo{address}{New York, NY, USA}, \bibinfo{pages}{664–687}.
\newblock
\showISBNx{9781450398930}
\urldef\tempurl%
\url{https://doi.org/10.1145/3563657.3595982}
\showDOI{\tempurl}


\bibitem[Hsieh et~al\mbox{.}(2024)]%
        {supporting}
\bibfield{author}{\bibinfo{person}{Jane Hsieh}, \bibinfo{person}{Angie Zhang}, \bibinfo{person}{Mialy Rasetarinera}, \bibinfo{person}{Erik Chou}, \bibinfo{person}{Daniel Ngo}, \bibinfo{person}{Jason Carpenter}, \bibinfo{person}{Karen Lightman}, \bibinfo{person}{Min~Kyung Lee}, {and} \bibinfo{person}{Haiyi Zhu}.} \bibinfo{year}{2024}\natexlab{}.
\newblock \showarticletitle{Supporting Gig Worker Needs and Advancing Policy Through Worker-Centered Data-Sharing}.
\newblock \bibinfo{journal}{\emph{CSCW '25}} (\bibinfo{year}{2024}).
\newblock
\newblock
\shownote{In submission}.


\bibitem[Kinder et~al\mbox{.}(2019)]%
        {actor_network}
\bibfield{author}{\bibinfo{person}{Eliscia Kinder}, \bibinfo{person}{Mohammad~Hossein Jarrahi}, {and} \bibinfo{person}{Will Sutherland}.} \bibinfo{year}{2019}\natexlab{}.
\newblock \showarticletitle{Gig Platforms, Tensions, Alliances and Ecosystems: An Actor-Network Perspective}.
\newblock \bibinfo{journal}{\emph{Proc. ACM Hum.-Comput. Interact.}} \bibinfo{volume}{3}, \bibinfo{number}{CSCW}, Article \bibinfo{articleno}{212} (\bibinfo{date}{nov} \bibinfo{year}{2019}), \bibinfo{numpages}{26}~pages.
\newblock
\urldef\tempurl%
\url{https://doi.org/10.1145/3359314}
\showDOI{\tempurl}


\bibitem[King et~al\mbox{.}(2023)]%
        {king2023privacy}
\bibfield{author}{\bibinfo{person}{Jennifer King}, \bibinfo{person}{Daniel Ho}, \bibinfo{person}{Arushi Gupta}, \bibinfo{person}{Victor Wu}, {and} \bibinfo{person}{Helen Webley-Brown}.} \bibinfo{year}{2023}\natexlab{}.
\newblock \showarticletitle{The Privacy-Bias Tradeoff: Data Minimization and Racial Disparity Assessments in U.S. Government}. In \bibinfo{booktitle}{\emph{Proceedings of the 2023 ACM Conference on Fairness, Accountability, and Transparency}} (Chicago, IL, USA) \emph{(\bibinfo{series}{FAccT '23})}. \bibinfo{publisher}{Association for Computing Machinery}, \bibinfo{address}{New York, NY, USA}, \bibinfo{pages}{492–505}.
\newblock
\showISBNx{9798400701924}
\urldef\tempurl%
\url{https://doi.org/10.1145/3593013.3594015}
\showDOI{\tempurl}


\bibitem[Kravchenko(2023)]%
        {fem_ethics}
\bibfield{author}{\bibinfo{person}{Elizaveta Kravchenko}.} \bibinfo{year}{2023}\natexlab{}.
\newblock \emph{\bibinfo{title}{The ethics of care and participatory design: a situated exploration}}.
\newblock \bibinfo{thesistype}{Ph.\,D. Dissertation}. \bibinfo{school}{The University of Texas at Austin}.
\newblock
\urldef\tempurl%
\url{https://doi.org/10.26153/tsw/48077}
\showDOI{\tempurl}


\bibitem[Lauren and Anandan(2024)]%
        {lauren12exploring}
\bibfield{author}{\bibinfo{person}{Rosario~Maria Lauren} {and} \bibinfo{person}{CR~Christi Anandan}.} \bibinfo{year}{2024}\natexlab{}.
\newblock \showarticletitle{Exploring the Challenges and Uncertainties faced by Gig Workers}.
\newblock \bibinfo{journal}{\emph{Journal of Academia and Industrial Research (JAIR)}} \bibinfo{volume}{12}, \bibinfo{number}{2} (\bibinfo{year}{2024}), \bibinfo{pages}{24--30}.
\newblock
\urldef\tempurl%
\url{https://doi.org/10.1177/0950017018785616}
\showDOI{\tempurl}


\bibitem[Mateescu and Nguyen(2019a)]%
        {explainer}
\bibfield{author}{\bibinfo{person}{Alexandra Mateescu} {and} \bibinfo{person}{A Nguyen}.} \bibinfo{year}{2019}\natexlab{a}.
\newblock \showarticletitle{Explainer}.
\newblock \bibinfo{journal}{\emph{Algorithmic management in the workplace}}  \bibinfo{volume}{6} (\bibinfo{year}{2019}), \bibinfo{pages}{1--15}.
\newblock
\urldef\tempurl%
\url{https://datasociety.net/wp-content/uploads/2019/02/DS_Algorithmic_Management_Explainer.pdf}
\showURL{%
\tempurl}


\bibitem[Mateescu and Nguyen(2019b)]%
        {monitoring_explainer}
\bibfield{author}{\bibinfo{person}{A Mateescu} {and} \bibinfo{person}{A Nguyen}.} \bibinfo{year}{2019}\natexlab{b}.
\newblock \bibinfo{title}{Explainer: Workplace Monitoring \& Surveillance.}
\newblock
\newblock
\urldef\tempurl%
\url{https://datasociety.net/wp-content/uploads/2019/02/DS_Workplace_Monitoring_Surveillance_Explainer.pdf}
\showURL{%
\tempurl}


\bibitem[Ming et~al\mbox{.}(2024)]%
        {homecarewages}
\bibfield{author}{\bibinfo{person}{Joy Ming}, \bibinfo{person}{Dana Gong}, \bibinfo{person}{Chit Sum~Eunice Ngai}, \bibinfo{person}{Madeline Sterling}, \bibinfo{person}{Aditya Vashistha}, {and} \bibinfo{person}{Nicola Dell}.} \bibinfo{year}{2024}\natexlab{}.
\newblock \showarticletitle{Wage Theft and Technology in the Home Care Context}.
\newblock \bibinfo{journal}{\emph{Proc. ACM Hum.-Comput. Interact.}} \bibinfo{volume}{8}, \bibinfo{number}{CSCW1}, Article \bibinfo{articleno}{151} (\bibinfo{date}{apr} \bibinfo{year}{2024}), \bibinfo{numpages}{30}~pages.
\newblock
\urldef\tempurl%
\url{https://doi.org/10.1145/3637428}
\showDOI{\tempurl}


\bibitem[Orr et~al\mbox{.}(2023)]%
        {necrocapitalism}
\bibfield{author}{\bibinfo{person}{Will Orr}, \bibinfo{person}{Kathryn Henne}, \bibinfo{person}{Ashlin Lee}, \bibinfo{person}{Jenna~Imad Harb}, {and} \bibinfo{person}{Franz~Carneiro Alphonso}.} \bibinfo{year}{2023}\natexlab{}.
\newblock \showarticletitle{Necrocapitalism in the gig economy: The case of platform food couriers in Australia}.
\newblock \bibinfo{journal}{\emph{Antipode}} \bibinfo{volume}{55}, \bibinfo{number}{1} (\bibinfo{year}{2023}), \bibinfo{pages}{200--221}.
\newblock
\urldef\tempurl%
\url{https://doi.org/10.1111/anti.12877}
\showDOI{\tempurl}


\bibitem[Purcell and Brook(2022)]%
        {boss}
\bibfield{author}{\bibinfo{person}{Christina Purcell} {and} \bibinfo{person}{Paul Brook}.} \bibinfo{year}{2022}\natexlab{}.
\newblock \showarticletitle{At least I’m my own boss! Explaining consent, coercion and resistance in platform work}.
\newblock \bibinfo{journal}{\emph{Work, Employment and Society}} \bibinfo{volume}{36}, \bibinfo{number}{3} (\bibinfo{year}{2022}), \bibinfo{pages}{391--406}.
\newblock
\urldef\tempurl%
\url{https://doi.org/10.1177/0950017020952661}
\showDOI{\tempurl}


\bibitem[Redden and Brand(2017)]%
        {redden2017data}
\bibfield{author}{\bibinfo{person}{Joanna Redden} {and} \bibinfo{person}{Jessica Brand}.} \bibinfo{year}{2017}\natexlab{}.
\newblock \showarticletitle{Data harm record}.
\newblock \bibinfo{journal}{\emph{Data Justice Lab}} (\bibinfo{year}{2017}).
\newblock
\urldef\tempurl%
\url{https://datajusticelab.org/data-harm-record/}
\showURL{%
\tempurl}


\bibitem[Sannon and Forte(2022)]%
        {sannon2022privacy}
\bibfield{author}{\bibinfo{person}{Shruti Sannon} {and} \bibinfo{person}{Andrea Forte}.} \bibinfo{year}{2022}\natexlab{}.
\newblock \showarticletitle{Privacy Research with Marginalized Groups: What We Know, What's Needed, and What's Next}.
\newblock \bibinfo{journal}{\emph{Proc. ACM Hum.-Comput. Interact.}} \bibinfo{volume}{6}, \bibinfo{number}{CSCW2}, Article \bibinfo{articleno}{455} (\bibinfo{date}{nov} \bibinfo{year}{2022}), \bibinfo{numpages}{33}~pages.
\newblock
\urldef\tempurl%
\url{https://doi.org/10.1145/3555556}
\showDOI{\tempurl}


\bibitem[Sannon et~al\mbox{.}(2022)]%
        {privacy}
\bibfield{author}{\bibinfo{person}{Shruti Sannon}, \bibinfo{person}{Billie Sun}, {and} \bibinfo{person}{Dan Cosley}.} \bibinfo{year}{2022}\natexlab{}.
\newblock \showarticletitle{Privacy, Surveillance, and Power in the Gig Economy}. In \bibinfo{booktitle}{\emph{Proceedings of the 2022 CHI Conference on Human Factors in Computing Systems}} (New Orleans, LA, USA) \emph{(\bibinfo{series}{CHI '22})}. \bibinfo{publisher}{Association for Computing Machinery}, \bibinfo{address}{New York, NY, USA}, Article \bibinfo{articleno}{619}, \bibinfo{numpages}{15}~pages.
\newblock
\showISBNx{9781450391573}
\urldef\tempurl%
\url{https://doi.org/10.1145/3491102.3502083}
\showDOI{\tempurl}


\bibitem[S\"{o}rries et~al\mbox{.}(2024)]%
        {sorries2024advocating}
\bibfield{author}{\bibinfo{person}{Peter S\"{o}rries}, \bibinfo{person}{David Leimst\"{a}dtner}, {and} \bibinfo{person}{Claudia M\"{u}ller-Birn}.} \bibinfo{year}{2024}\natexlab{}.
\newblock \showarticletitle{Advocating Values through Meaningful Participation: Introducing a Method to Elicit and Analyze Values for Enriching Data Donation Practices in Healthcare}.
\newblock \bibinfo{journal}{\emph{Proc. ACM Hum.-Comput. Interact.}} \bibinfo{volume}{8}, \bibinfo{number}{CSCW1}, Article \bibinfo{articleno}{16} (\bibinfo{date}{apr} \bibinfo{year}{2024}), \bibinfo{numpages}{32}~pages.
\newblock
\urldef\tempurl%
\url{https://doi.org/10.1145/3637293}
\showDOI{\tempurl}


\bibitem[Stein et~al\mbox{.}(2023)]%
        {uuapp}
\bibfield{author}{\bibinfo{person}{Jake M~L Stein}, \bibinfo{person}{Vidminas Vizgirda}, \bibinfo{person}{Max Van~Kleek}, \bibinfo{person}{Reuben Binns}, \bibinfo{person}{Jun Zhao}, \bibinfo{person}{Rui Zhao}, \bibinfo{person}{Naman Goel}, \bibinfo{person}{George Chalhoub}, \bibinfo{person}{Wael~S Albayaydh}, {and} \bibinfo{person}{Nigel Shadbolt}.} \bibinfo{year}{2023}\natexlab{}.
\newblock \showarticletitle{‘You are you and the app. There’s nobody else.’: Building Worker-Designed Data Institutions within Platform Hegemony}. In \bibinfo{booktitle}{\emph{Proceedings of the 2023 CHI Conference on Human Factors in Computing Systems}} (Hamburg, Germany,) \emph{(\bibinfo{series}{CHI '23})}. \bibinfo{publisher}{Association for Computing Machinery}, \bibinfo{address}{New York, NY, USA}, Article \bibinfo{articleno}{281}, \bibinfo{numpages}{26}~pages.
\newblock
\showISBNx{9781450394215}
\urldef\tempurl%
\url{https://doi.org/10.1145/3544548.3581114}
\showDOI{\tempurl}


\bibitem[Stewart and Stanford(2017)]%
        {australia_regulation}
\bibfield{author}{\bibinfo{person}{Andrew Stewart} {and} \bibinfo{person}{Jim Stanford}.} \bibinfo{year}{2017}\natexlab{}.
\newblock \showarticletitle{Regulating work in the gig economy: What are the options?}
\newblock \bibinfo{journal}{\emph{The Economic and Labour Relations Review}} \bibinfo{volume}{28}, \bibinfo{number}{3} (\bibinfo{year}{2017}), \bibinfo{pages}{420--437}.
\newblock
\urldef\tempurl%
\url{https://doi.org/10.1177/1035304617722461}
\showDOI{\tempurl}


\bibitem[Vincent et~al\mbox{.}(2019)]%
        {strikes}
\bibfield{author}{\bibinfo{person}{Nicholas Vincent}, \bibinfo{person}{Brent Hecht}, {and} \bibinfo{person}{Shilad Sen}.} \bibinfo{year}{2019}\natexlab{}.
\newblock \showarticletitle{“Data Strikes”: Evaluating the Effectiveness of a New Form of Collective Action Against Technology Companies}. In \bibinfo{booktitle}{\emph{The World Wide Web Conference}} (San Francisco, CA, USA) \emph{(\bibinfo{series}{WWW '19})}. \bibinfo{publisher}{Association for Computing Machinery}, \bibinfo{address}{New York, NY, USA}, \bibinfo{pages}{1931–1943}.
\newblock
\showISBNx{9781450366748}
\urldef\tempurl%
\url{https://doi.org/10.1145/3308558.3313742}
\showDOI{\tempurl}


\bibitem[Vincent et~al\mbox{.}(2021)]%
        {leverage}
\bibfield{author}{\bibinfo{person}{Nicholas Vincent}, \bibinfo{person}{Hanlin Li}, \bibinfo{person}{Nicole Tilly}, \bibinfo{person}{Stevie Chancellor}, {and} \bibinfo{person}{Brent Hecht}.} \bibinfo{year}{2021}\natexlab{}.
\newblock \showarticletitle{Data Leverage: A Framework for Empowering the Public in its Relationship with Technology Companies}.
\newblock \bibinfo{journal}{\emph{Proceedings of the 2021 ACM Conference on Fairness, Accountability, and Transparency}} (\bibinfo{year}{2021}), \bibinfo{pages}{215–227}.
\newblock
\urldef\tempurl%
\url{https://doi.org/10.1145/3442188.3445885}
\showDOI{\tempurl}


\bibitem[Wood et~al\mbox{.}(2019a)]%
        {goodgig}
\bibfield{author}{\bibinfo{person}{Alex~J Wood}, \bibinfo{person}{Mark Graham}, \bibinfo{person}{Vili Lehdonvirta}, {and} \bibinfo{person}{Isis Hjorth}.} \bibinfo{year}{2019}\natexlab{a}.
\newblock \showarticletitle{Good Gig, Bad Gig: Autonomy and Algorithmic Control in the Global Gig Economy}.
\newblock \bibinfo{journal}{\emph{Work, Employment and Society}} \bibinfo{volume}{33}, \bibinfo{number}{1} (\bibinfo{year}{2019}), \bibinfo{pages}{56--75}.
\newblock
\urldef\tempurl%
\url{https://doi.org/10.1177/0950017018785616}
\showDOI{\tempurl}


\bibitem[Wood et~al\mbox{.}(2019b)]%
        {wood2019networked}
\bibfield{author}{\bibinfo{person}{Alex~J Wood}, \bibinfo{person}{Mark Graham}, \bibinfo{person}{Vili Lehdonvirta}, {and} \bibinfo{person}{Isis Hjorth}.} \bibinfo{year}{2019}\natexlab{b}.
\newblock \showarticletitle{Networked but commodified: The (dis) embeddedness of digital labour in the gig economy}.
\newblock \bibinfo{journal}{\emph{Sociology}} \bibinfo{volume}{53}, \bibinfo{number}{5} (\bibinfo{year}{2019}), \bibinfo{pages}{931--950}.
\newblock
\urldef\tempurl%
\url{https://doi.org/10.1177/0038038519828}
\showDOI{\tempurl}


\bibitem[Yao et~al\mbox{.}(2021)]%
        {atomized}
\bibfield{author}{\bibinfo{person}{Zheng Yao}, \bibinfo{person}{Silas Weden}, \bibinfo{person}{Lea Emerlyn}, \bibinfo{person}{Haiyi Zhu}, {and} \bibinfo{person}{Robert~E. Kraut}.} \bibinfo{year}{2021}\natexlab{}.
\newblock \showarticletitle{Together But Alone: Atomization and Peer Support among Gig Workers}.
\newblock \bibinfo{journal}{\emph{Proc. ACM Hum.-Comput. Interact.}} \bibinfo{volume}{5}, \bibinfo{number}{CSCW2}, Article \bibinfo{articleno}{391} (\bibinfo{date}{oct} \bibinfo{year}{2021}), \bibinfo{numpages}{29}~pages.
\newblock
\urldef\tempurl%
\url{https://doi.org/10.1145/3479535}
\showDOI{\tempurl}


\bibitem[Zalnieriute(2021)]%
        {zalnieriute2021transparency}
\bibfield{author}{\bibinfo{person}{Monika Zalnieriute}.} \bibinfo{year}{2021}\natexlab{}.
\newblock \showarticletitle{"Transparency Washing" in the Digital Age: A Corporate Agenda of Procedural Fetishism}.
\newblock \bibinfo{journal}{\emph{Critical Analysis L.}}  \bibinfo{volume}{8} (\bibinfo{year}{2021}), \bibinfo{pages}{139}.
\newblock
\urldef\tempurl%
\url{https://doi.org/10.33137/cal.v8i1.36284}
\showDOI{\tempurl}


\bibitem[Zhang et~al\mbox{.}(2023)]%
        {probes}
\bibfield{author}{\bibinfo{person}{Angie Zhang}, \bibinfo{person}{Alexander Boltz}, \bibinfo{person}{Jonathan Lynn}, \bibinfo{person}{Chun-Wei Wang}, {and} \bibinfo{person}{Min~Kyung Lee}.} \bibinfo{year}{2023}\natexlab{}.
\newblock \showarticletitle{Stakeholder-Centered AI Design: Co-Designing Worker Tools with Gig Workers through Data Probes}. In \bibinfo{booktitle}{\emph{Proceedings of the 2023 CHI Conference on Human Factors in Computing Systems}} (Hamburg, Germany) \emph{(\bibinfo{series}{CHI '23})}. \bibinfo{publisher}{Association for Computing Machinery}, \bibinfo{address}{New York, NY, USA}, Article \bibinfo{articleno}{859}, \bibinfo{numpages}{19}~pages.
\newblock
\showISBNx{9781450394215}
\urldef\tempurl%
\url{https://doi.org/10.1145/3544548.3581354}
\showDOI{\tempurl}


\bibitem[Zhang et~al\mbox{.}(2022)]%
        {imaginaries}
\bibfield{author}{\bibinfo{person}{Angie Zhang}, \bibinfo{person}{Alexander Boltz}, \bibinfo{person}{Chun~Wei Wang}, {and} \bibinfo{person}{Min~Kyung Lee}.} \bibinfo{year}{2022}\natexlab{}.
\newblock \showarticletitle{Algorithmic Management Reimagined For Workers and By Workers: Centering Worker Well-Being in Gig Work}. In \bibinfo{booktitle}{\emph{Proceedings of the 2022 CHI Conference on Human Factors in Computing Systems}} (New Orleans, LA, USA) \emph{(\bibinfo{series}{CHI '22})}. \bibinfo{publisher}{Association for Computing Machinery}, \bibinfo{address}{New York, NY, USA}, Article \bibinfo{articleno}{14}, \bibinfo{numpages}{20}~pages.
\newblock
\showISBNx{9781450391573}
\urldef\tempurl%
\url{https://doi.org/10.1145/3491102.3501866}
\showDOI{\tempurl}


\bibitem[Zong and Matias(2024)]%
        {refusal}
\bibfield{author}{\bibinfo{person}{Jonathan Zong} {and} \bibinfo{person}{J.~Nathan Matias}.} \bibinfo{year}{2024}\natexlab{}.
\newblock \showarticletitle{Data Refusal from Below: A Framework for Understanding, Evaluating, and Envisioning Refusal as Design}.
\newblock \bibinfo{journal}{\emph{ACM J. Responsib. Comput.}} \bibinfo{volume}{1}, \bibinfo{number}{1}, Article \bibinfo{articleno}{10} (\bibinfo{date}{mar} \bibinfo{year}{2024}), \bibinfo{numpages}{23}~pages.
\newblock
\urldef\tempurl%
\url{https://doi.org/10.1145/3630107}
\showDOI{\tempurl}


\end{thebibliography}

%%
%% If your work has an appendix, this is the place to put it.
\appendix

\end{document}